\documentclass[12pt]{article}
\usepackage{cite, bm}
\def\be{\begin{equation}}
\def\ee{\end{equation}}
\def\bea{\begin{eqnarray}}
\def\eea{\end{eqnarray}}

\begin{document}

\begin{center}
{\Large{\bf Scaling Behavior of Fractional Growth Processes}}

\vskip .5cm {\large Reza Torabi}
\vskip .1cm
{\it Department of Physics and Astronomy, University of Calgary, Calgary, Alberta T2N 1N4, Canada}\\
{\sl e-mail: reza.torabi@ucalgary.ca}\\
\end{center}

\begin{abstract}

An exact solution is introduced for one dimensional
space-fractional Edwards-Wilkinson equation. It is
shown that the roughness obeys the Family-Viscek dynamic scaling form
and the scaling exponents is derived. It is seen that
the scaling exponents are different from those coming from ordinary
Edwards-Wilkinson equation and depend on the fractional order. Scaling exponents in the presence of correlated noise is also obtained.

\end{abstract}

\vskip .5cm
\newpage

\section{Introduction}

Fractional calculus \cite{Podlubny,Oldham,Miller,Samko} deals with
the generalization of differentiation and integration to fractional
order. It is a powerful mathematical tool that is used to describe
complex systems in which conventional approaches have failed. The
interest in fractional calculus has grown rapidly in recent years
due to its various applications.

There are many different areas where fractional equations describe
real processes such as anomalous diffusion
\cite{Metzler,Weissman,Shlesinger,Leith}, chaotic dynamics
\cite{Zaslavsky,Zaslavsky2}, dynamics in complex media
\cite{Tarasov1,Tarasov2,Tarasov3,Tarasov4,Tarasov5}, relaxation in
complex viscoelastic materials \cite{Giona}, random processes of
Levy type \cite{Saichev,Uchaikin,Meerschaert}, heat transfer in
fractal media \cite{Povstenko,Hristov}, acoustic waves in complex
media \cite{Holm}, non-Hamiltonian mechanics
\cite{Tarasov6,Tarasov7}, theory of long-range interactions
\cite{Laskin,Tarasov8}, fractional Fokker-Planck equation
\cite{Ervin}, electromagnetic field of fractal distribution of
charged particles \cite{Tarasov9}, electromagnetic field of fractal
media \cite{Baskin,Tarasov10}, damped oscillations \cite{Ryabov},
Navier-Stokes equations on Cantor sets \cite{Yang}, fractional
schr\"{o}dinger equation \cite{Laskin2} and generalized Langevin
equation \cite{Taloni1,Lizana,Taloni2}.

In fact, the generalization of many equations of physics to
fractional order has been carried out due to various reasons such as
the presence of memory effects, dissipation, intermittency, fractal
media, long-range interactions and etc. One of these reasons that we
are interested in is the presence of fractal media. Fractional
calculus allow us to take into account the fractal geometry of the
medium. In this case the fractal structure is disregarded and
instead, the process is described by a fractional equation
\cite{Prehl,Carpinteri,Chen,Metzler}. Here we consider the growth
process of an interface in a porous (disordered) medium . Since
porous media can be considered as fractal structures, we use
fractional calculus to describe the process.

The growth process of interfaces far from equilibrium, where the
height gradient is negligible, is governed by Edwards-Wilkinson
equation. The dynamics of interface in disordered media is described
by adding a quenched noise to the Edwards-Wilkinson equation. It is
hard to study this equation due to the presence of the quenched
noise. Here we present an analytical approach based on the
fractional calculus to find how the exponents change in disordered
media and to see if roughness obeys the dynamic scaling form in
these media.

The paper is organized as follows: In section II, a brief
Introduction to the Edwards-Wilkinson equation both in ordinary and
disordered media is presented. In section III, we introduce a
fractional Edwards-Wilkinson equation for interface in disordered
media and solve it analytically. We show that the roughness of an
interface in a disordered medium obeys the Family-Viscek scaling
form and find the scaling exponents. In section IV, the
height-height correlation function is discussed. The fractional
Edwards-Wilkinson equation in the presence of correlated noise is
also studied in section V. Finally the conclusion is presented in
section VI.

\section{Interfaces in disordered media}

The dynamics of an interface, where $|\nabla h|\ll1$, is governed by
the Edwards-Wilkinson equation. This equation is a linear Langevin
equation with the form \cite{Barabasi,Nattermann}
\begin{equation}
	\frac{\partial h({\bf x},t)}{\partial t}=\nu \nabla^2
	h({\bf x},t)+\eta({\bf x},t),
\end{equation}
where $h({\bf x},t)$ is the height of a $d$ dimensional interface at
point ${\bf x}$ and time $t$. The first term on the right hand side
describes relaxation of the interface and $\nu$ is sometimes called
surface tension. $\eta({\bf x},t)$ is a stochastic noise which
reflects the random fluctuations. It has zero average, $\langle
\eta({\bf x},t)\rangle=0$, and its second momentum is given by
\begin{equation}
	\langle \eta({\bf x},t)\eta({\bf x'},t') \rangle=2D\delta^d({\bf
		x}-{\bf x'})\delta(t-t').
\end{equation}
Here $\langle\rangle$ denotes average over the noise distribution
and Eq. (2) implies that the noise has no correlation in space and
time. The interface width, $W$, which characterizes the roughness of
the interface is defined by the root mean square fluctuation in the
height.

Because of the linear character of the Edwards-Wilkinson equation,
one can solve it straightforwardly and show that the width of the
interface obeys the Family-Viscek dynamic scaling form
\begin{equation}
	W^2(L,t)\simeq L^{2\alpha} F(\frac{t}{L^z}),
\end{equation}
and also obtain the scaling exponents. $\alpha$ and $z$ are known as
roughness and dynamic exponents, respectively. $F(u)$ is a scaling
function with the properties $F(u)\rightarrow const$ as
$u\rightarrow\infty$ and $F(u)\rightarrow u^{(2-d)/2}$ as
$u\rightarrow0$. For $t\ll L^z$ the $L$ dependence drops out and (3)
yields to $W^2(L,t)\simeq t^{2\beta}$ with $\beta=\alpha/z$. $\beta$
is growth exponent and the scaling exponents of the
Edwards-Wilkinson equation are
\begin{equation}
	\alpha=\frac{2-d}{2},\;\;\; \beta=\frac{2-d}{4},\;\;\;z=2.
\end{equation}
The discrete version of the model, governs by (1), is "random
deposition with surface relaxation". Surface relaxations allow the
deposited particles to diffuse along the interface up to a finite
distance to find the position with lowest height.

In the case of interface in a disordered medium, a quenched noise,
that is generated by the disorder is added to the Edward-Wilkinson
equation as
\begin{equation}
	\frac{\partial h({\bf x},t)}{\partial t}=\nu \nabla^2 h({\bf
		x},t)+\eta({\bf x},t)+\eta({\bf x},h),
\end{equation}
It is  assumed that $\eta({\bf x},h)$ has zero mean, $\langle
\eta({\bf x},h)\rangle=0$, and the correlation of the form
\[
\langle \eta({\bf x},h)\eta({\bf x'},h') \rangle=\delta^d({\bf
	x}-{\bf x'})\Delta(h-h'),
\]
where $\Delta$ is a function that falls of rapidly when its
arguments is large. Equation (5) can not be solved in a
straightforward manner due to the nature of the quenched noise. To
understand the effect of quenched noise, there has been a lots of
numerical studies \cite{Barabasi}. In general it is agreed that
quenched noise produce anomalous roughening with a roughness
exponent larger than the value produced by (4).

\section{The fractional Edward-Wilkinson equation}

In disordered media the deposited particle undergoes anomalous
diffusion, due to the effect of medium, to find the position with
lowest height. This yields to an anomalous relaxation process.
According to this fact, we introduce a fractional Edward-Wilkinson
equation to describe the process. In other words, to incorporate the
effect of disordered media, we generalize Edward-Wilkinson equation
to fractional order instead of adding the quenched noise. This
enable us to solve the problem via analytical approaches. The
generalized Edward-Wilkinson equation is in the form

\begin{equation}
	\frac{\partial h({{\bf x}},t)}{\partial t}=\nu_{\gamma}
	\frac{\partial^\gamma}{\partial |{\bf x}|^\gamma} h({{\bf
			x}},t)+\eta({{\bf x}},t),
\end{equation}
where $\gamma$ is the fractional order and $\gamma<2$ due to the
porosity. $\nu_\gamma$ is a generalized surface tension and the
noise is again a white one. The fractional derivative is defined via
its Fourier transform according to \cite{Saichev,Taloni1,Taloni3}
\[
F_{\bf k} \left\{\frac{\partial^\gamma}{\partial |{\bf x}|^\gamma}
\right\}\equiv -|{\bf k}|^\gamma.
\]
Another definition for the above fractional derivative is given in
terms of the Laplacian as ${\partial^\gamma}/{\partial |{\bf
		x}|^\gamma}=-(-\nabla^2)^{\gamma/2}$ \cite{Samko}. In fact, this
equation is an special case of the generalized elastic model
proposed by \cite{Taloni1,Taloni3}. It can be achieved by
considering the hydrodynamic friction kernel to be local and
proportional to the Dirac delta function. As we need, the authors of
\cite{Taloni1,Taloni3} showed that, in the systems obeying (6), the
particle placed at ${\bf x}$ undergoes a subdiffusive motion (a
fractional Brownian motion) due to the spatial interactions with the
other system components.

Equation (6) can be solved in a straight forward manner. Fourier
transform of (6) in space and time gives
\begin{equation}
	h({\bf k},\omega)=\frac{1}{\nu_{\gamma} k^\gamma-i\omega} \eta({\bf
		k},\omega),
\end{equation}
where $k=|{\bf k}|$ and
\[
h({{\bf k}},\omega)=\int d^d{\bf x}\;dt\; e^{-i({\bf k}\cdot {\bf x}-\omega t)}h({{\bf
		x}},t),
\]
\[
\eta({{\bf k}},\omega)=\int d^d{\bf x}\;dt\; e^{-i({\bf k}\cdot {\bf x}-\omega t)}\eta({{\bf
		x}},t).
\]

In order to describe a growth process that is started from a flat
substrate at $t=0$, we set $\eta({\bf x},t)=0$ for $t<0$. The second
momentum of the noise in Fourier space then takes the form
\begin{equation}
	\langle \eta({\bf k},\omega)\eta({\bf k'},\omega')
	\rangle=2D(2\pi)^d\delta^d({\bf k}+{\bf k'})\int_0^\infty d\tau
	e^{i(\omega+\omega')\tau}.
\end{equation}

The width of an interface of size $L$ is
\[
W^2(L,t)=\left\langle \frac{1}{N} \sum_{\bf x} \left( h({\bf
	x},t)-\frac{1}{N}\sum_{\bf x'} h({\bf x'},t) \right)^2
\right\rangle,
\]
where it can be expressed as \cite{Nattermann}
\begin{equation}
	W^2(L,t)=\int \frac{d^d {\bf k}}{(2\pi)^d}\frac{d^d {\bf
			k'}}{(2\pi)^d}\langle h({\bf k},t)h({\bf k'},t) \rangle.
\end{equation}
Thus, in order to find $W^2(L,t)$, we need to calculate the quantity
$\langle h({\bf k},t)h({\bf k'},t) \rangle$. Using (7), (8) in (9)
and performing the $\omega$ integral, we obtain
\begin{equation}
	\langle h({\bf k},t)h({\bf k'},t) \rangle=\frac{D}{\nu_\gamma
		k^\gamma}(1-e^{-2\nu_\gamma k^\gamma t}){(2\pi)^d}\delta^d({\bf
		k}+{\bf k'}),
\end{equation}
where we have used the identity
\[
\int_{-\infty}^{+\infty}\frac{e^{im x}}{n-ix}=-2\pi e^{mn}.
\]
We note that the relaxation time $\tau$ of modes of wave vector
${\bf k}$ is now $\tau\simeq {1}/{\nu_\gamma k^\gamma}$ instead of
$\tau\simeq {1}/{\nu k^2}$.

Putting (10) in (9) and carrying out the integral using spherical
coordinates yields
\begin{equation}
	W^2(L,t)=
	K_d(\frac{D}{\nu_\gamma})\int_{\frac{2\pi}{L}}^{\frac{\pi}{a}}
	\frac{dk}{k^{\gamma+1-d}} (1-e^{-2\nu_\gamma k^\gamma t}),
\end{equation}
where $K_d^{-1}=2^{d-1}\pi^{d/2}\Gamma(d/2)$ and $a$ is the lattice
spacing. Performing the integration by parts, (11) will cast in the
Family-Viscek scaling form
\begin{equation}
	W^2(L,t)=A+\frac{D}{\nu_\gamma} L^{\gamma-d} f_d(\frac{\nu_\gamma
		t}{L^\gamma})+O(e^{-\frac{2\pi\gamma\nu_\gamma t}{a^\gamma}}),
\end{equation}
where
\[
A=\frac{K_d}{d-\gamma}(\frac{D}{\nu_\gamma})(\frac{\pi}{a})^{d-\gamma},
\]
and
\[
f_d(x)=\frac{K_d}{(\gamma-d)(2\pi)^{\gamma-d}}{\bigg(}1-e^{-2(2\pi)^\gamma
	x}+[{2(2\pi)^\gamma x}]^{1-\frac{d}{\gamma}}\int_{2(2\pi)^\gamma
	x}^\infty y^{\frac{d}{\gamma}-1}e^{-y}dy {\bigg)}.
\]
At large $x$, $f_d(x)$ approaches a constant and for $x\ll1$ we have
$f(x)\propto x^{\frac{\gamma-d}{\gamma}}$,

The scaling exponents can be easily find by comparing (12) with the
family-viscek scaling form (3) as
\begin{equation}
	\alpha=\frac{\gamma-d}{2},\;\;\;
	\beta=\frac{\gamma-d}{2\gamma},\;\;\;z=\gamma,
\end{equation}
where $z=\alpha/\beta$. The fractional order has affected the
scaling exponents through the relation (13).

The fractional order $\gamma=2$ corresponds to the ordinary
Edwards-Wilkinson equation. For fractional order $\gamma>2$ all the
scaling exponents take larger values than the ones coming from the
ordinary Edwards-Wilkinson equation. However, all the exponents have
smaller values in the case of $\gamma<2$.

\section{Height-height correlation function}

Height-height correlation function is another quantity of interest
which can be used to determine scaling exponents. This quantity can
be measured experimentally. It has the form
\begin{equation}
	C({\bf r},t)\equiv\langle |h({\bf r},t)-({\bf 0},t)|^2 \rangle.
\end{equation}
which is written as
\begin{equation}
	C({\bf r},t)=\int_{\frac{2\pi}{L}}^{\frac{\pi}{a}} \frac{d^d{\bf
			k}}{(2\pi)^d}\frac{2D}{\nu_\gamma k^\gamma} (1-e^{-2\nu_\gamma
		k^\gamma t})[1-cos({\bf k}\cdot {\bf r})],
\end{equation}
in momentum space. Using spherical coordinate and performing the angular integration yields
\begin{equation}
	C({\bf r},t)=\frac{2D}{\nu_\gamma}K_d\int_{0}^{\frac{\pi}{a}} dk
	k^{d-1-\gamma} (1-e^{-2\nu_\gamma k^\gamma
		t})\left[1-\Gamma\left(\frac{d}{2}\right)J_{\frac{d}{2}-1}(kr)\left(\frac{kr}{2}\right)^{1-\frac{d}{2}}\right],
\end{equation}
for the case of $L\rightarrow\infty$. $r=|{\bf r}|$ and $J_n(x)$ is
the Bessel function of order $n$. We are interested in finding the
behavior of height-height correlation function at large $r$. To this
end, we use change of variables $y=kr$ and $x={\nu_\gamma
	t}/{r^\gamma}$. The height-height correlation function takes the
form
\begin{equation}
	C({\bf r},t)=\frac{2D}{\nu_\gamma} r^{\gamma-d} f_d(\frac{\nu_\gamma
		t}{r^\gamma})+O(1),
\end{equation}
is the scaling function. Equation (17) states that the height-height
correlation function also has a universal scaling behavior in the
presence of porosity. 

\section{Correlated noise}

In the previous sections we assumed that the noise is an
uncorrelated one, for instance, with a gaussian distribution. But it
is not the only possible type of noise in physical systems. In this section we
are going to study the growth of an interface that is governed by a
generalized Edwards-Wilkinson equation in the presence of spatially
and temporally correlated noise using the scaling arguments.

In the case that the noise has long-range spatial and temporal
correlations, we have
\begin{equation}
	\langle \eta({\bf x},t)\eta({\bf x'},t') \rangle\sim|{\bf x}-{\bf
		x'}|^{2\psi-d}|t-t'|^{2\phi-1}.
\end{equation}
where $\psi(\phi)$ characterize the decay of spatial (temporal)
correlations. The case of uncorrelated noise corresponds to $\psi=0$
and $\phi=0$. To see how the correlated noise influence the
growth exponents, we use scaling arguments.

If the interface is self-affine, then the dynamic equation (6) would
not change by the rescaling
\begin{equation}
	{\bf x}\rightarrow b{\bf x}, \;\;\; h\rightarrow b^\alpha h, \;\;\;
	t\rightarrow t^z
\end{equation}
The noise second momentum rescales as
\[
\langle \eta(b{\bf x},b^z t)\eta(b{\bf x'},b^z t') \rangle\sim
b^{2\psi-d+2z\phi-z} |{\bf x}-{\bf x'}|^{2\psi-d}|t-t'|^{2\phi-1},
\]
which indicates that $\eta\rightarrow
b^{\psi-\frac{d}{2}+z\phi-\frac{z}{2}}\eta$. Thus, after rescaling,
the dynamic equation (6) becomes
\begin{equation}
	\frac{\partial h({\bf x},t)}{\partial t}=b^{z-\gamma}\nu_{\gamma}
	\frac{\partial^\gamma}{\partial |{\bf x}|^\gamma} h({{\bf x}},t)+
	b^{\psi-\frac{d}{2}+z\phi+\frac{z}{2}-\alpha} \eta({\bf x},t).
\end{equation}
Since the dynamic equation (6) is invariant under the scaling
transformation (19), the exponents will be
\begin{equation}
	\alpha=\psi+\gamma\phi+\frac{\gamma-d}{2},\;\;\;
	\beta=\frac{\psi}{\gamma}+\phi+\frac{\gamma-d}{2\gamma},\;\;\;z=\gamma.
\end{equation}
As one can see the presence of correlated noise change the scaling
exponent. Note that, in the case of uncorrelated noise we find the
exponents of (13). Equation (21) reflect the fact that the correlated
noise increase the roughness exponent.

\section{Conclusion}

We considered a one dimensional space-fractional Edwards-Wilkinson
equation and solved it exactly. It has been shown that the roughness obeys
Family-Viscek dynamic scaling form in the case of this generalized
Edwards-Wilkinson equation in similar to the ordinary
Edwards-Wilkinson equation. We obtained the scaling exponents and
showed that the fractional order affects the these exponents. Also, the scaling exponents in the presence of correlated noise has been obtained.

This equation is a linear Langevin equation and is employed in many
areas of physics such as fluid flow in porous media, flux lines in
superconductors, deposition processes, bacterial growth and so on.
Therefore, generalization of ordinary Edwards-Wilkinson equation to
a fractional equation can be interesting due to its application not
only in surface growth phenomena but in many areas of physics.


\begin{thebibliography}{widest-label}
	
	\bibitem{Podlubny} I. Podlubny, {\it Fractional Differential Equations}, Academic Press, San Diego (1999).
	\bibitem{Oldham} K. B. Oldham and J. Spanier, {\it The Fractional Calculus}, Academic Press, New York (1974).
	\bibitem{Miller} K. S. Miller and B. Ross, {\it An Introduction to the Fractional Calculus and Fractional Differential Equations}, Wiley-Interscience (1993).
	\bibitem{Samko} S. G. Samko, A. A. Kilbas and O. I. Marichev, {\it Fractional Integrals and Derivatives}, Gordon and Breach Science Publishers, Amsterdam (1993).
	\bibitem{Metzler} R. Metzler and J. Klafter, Phys. Rep. {\bf 339}, 1 (2000).
	\bibitem{Weissman} M. Weissman, Rev. Mod. Phys. {\bf 60}, 537 (1988).
	\bibitem{Shlesinger} M. Shlesinger, Annu. Rev. Phys. Chem. {\bf 39}, 269 (1988).
	\bibitem{Leith} J. R. Leith, Signal Processing {\bf 83} 2397 (2003).
	\bibitem{Zaslavsky} G. M. Zaslavsky, Phys. Rep. {\bf 371}, 461 (2002).
	\bibitem{Zaslavsky2} G. M. Zaslavsky, {\it Hamiltonian Chaos and Fractional Dynamics}, Oxford Univiversity Press (2005).
	\bibitem{Tarasov1} V. E. Tarasov, Phys. Lett. A {\bf 336}, 167 (2005).
	\bibitem{Tarasov2} V. E. Tarasov, Ann. Phys. {\bf 318}, 286 (2005).
	\bibitem{Tarasov3} V. E. Tarasov, Chaos {\bf 15}, 023102 (2005).
	\bibitem{Tarasov4} V. E. Tarasov, Phys. Lett. A {\bf 341}, 467 (2005).
	\bibitem{Tarasov5} V. E. Tarasov, Mod. Phys. Lett. B {\bf 19}, 721 (2005).
	\bibitem{Giona} M, Giona, S. Cerbelli and H. E. Roman, Physica A {\bf 191}, 449 (1992).
	\bibitem{Saichev} A. I. Saichev, G. M. Zaslavsky, Chaos {\bf 7}, 753 (1997).
	\bibitem{Uchaikin} V. V. Uchaikin, J. Exper. Theor. Phys. {\bf 97}, 810 (2003).
	\bibitem{Meerschaert} M. M. Meerschaert, D. A. Benson, B. Baeumer, Phys. Rev. E {\bf 63}, 021112 (2001).
	\bibitem{Povstenko} V. Z. Povstenko, J. Therm. Stress. {\bf 28}, 83 (2005).
	\bibitem{Hristov} J. Hristov, Eur. Phys. J. Spec. {\bf 193}, 229 (2011).
	\bibitem{Holm} S. Holm and S. P. Näsholm, J. Acoust. Soc. Am. {\bf 130}, 2195 (2011).
	\bibitem{Tarasov6} V. E. Tarasov, Phys. Rev. E {\bf 71}, 011102 (2005);
	\bibitem{Tarasov7} V. E. Tarasov, J. Phys. A {\bf 38}, 5929 (2005).
	\bibitem{Laskin} N. Laskin and G. M. Zaslavsky, Physica A {\bf 368}, 38 (2005).
	\bibitem{Tarasov8} V. E. Tarasov and G. M. Zaslavsky, Chaos {\bf 16}, 023110 (2006).
	\bibitem{Ervin} V. G. Ervin and J. P. Roop, Numer. Meth. PDE {\bf 22}, 558 (2006).
	\bibitem{Tarasov9} V. E. Tarasov, Phys. Plasmas {\bf 12}, 082106 (2005).
	\bibitem{Baskin} E. Baskin and A. Iomin, Chaos Solitons Fract. {\bf 44}, 335 (2011).
	\bibitem{Tarasov10} V. E. Tarasov, Mod. Phys. Lett. A {\bf 21}, 1587 (2006).
	\bibitem{Ryabov} Ya. E. Ryabov and A. Puzenko, Phys. Rev. B {\bf 66}, 184201 (2002).
	\bibitem{Yang} X. J. Yang, D. Baleanu and J. A. T. Machado, Math. Probl. Eng. {\bf 2013}, 1 (2013).
	\bibitem{Laskin2} N. Laskin, Phys. Rev. E {\bf 66}, 056108 (2002).
	\bibitem{Taloni1} A. Taloni, A. Chechkin, J. Klafter, Phys. Rev. E {\bf 82}, 061104 (2010).
	\bibitem{Lizana} L. Lizana, T. Ambj\"{o}rnsson, A. Taloni, E. Barkai and M. A. Lomholt, Phys. Rev. E {\bf 81}, 051118 (2010).
	\bibitem{Taloni2} A. Taloni, A. Chechkin and J. Klafter, Math. Model. Nat. Phenom. {\bf 8}, 2 (2013).
	\bibitem{Prehl} J. Prehl, {\it Diffusion on fractals and space-fractional diffusion equations}, M. Sc Thesis, Chemnitz university of thechnology (2010).
	\bibitem{Carpinteri} A. Carpinteri and A. Sapora, J. Appl .Math. Mech. {\bf 90}, 203 (2010).
	\bibitem{Chen} W. Chen, H. G. Sun, X. Zhang and D. Korosak, Comput. Math. Appl. {\bf 59}, 1754 (2010).
	\bibitem{Barabasi} A. L. Barabasi and H. E. Stanley, {\it Fractal Concepts in Surface Growth}, Cambridge Univiversity Press (1995).
	\bibitem{Nattermann} T. Nattermann and L. H. Tang, Phys. Rev. A {\bf 45}, 7156 (1992).
	\bibitem{Taloni3} A. Taloni, A. Chechkin, J. Klafter, Phys. Rev. Lett. {\bf 104}, 160602 (2010).
	
	
\end{thebibliography}
\end{document}